**Optical study of laser biospeckle activity in leaves of *Jatropha curcas* L. A noninvasive analysis of foliar endophyte colonization**

D'Jonsiles, Maria Fernanda[a,b]; Galizzi, Gustavo Ernesto[c,d]; Dolinko, Andres Ezequiel[a,b]; Novas, María Victoria[a,b]; Ceriani Nakamurakare, Esteban [a,b]; Carmarán, Cecilia Cristina[a,b].

[a] Universidad de Buenos Aires, Facultad de Ciencias Exactas y Naturales, Departamento de Biodiversidad y Biología Experimental, Intendente Güiraldes 2160 - Ciudad Universitaria (CP1428EHA), Ciudad Autónoma de Buenos Aires, Argentina.

[b]CONICET, Universidad de Buenos Aires, Instituto de Micología y Botánica (INMIBO). Intendente Güiraldes 2160 - Ciudad Universitaria, Ciudad Autónoma de Buenos Aires, Argentina.

[c]Instituto de Física Rosario (CONICET-UNR), Blvd. 27 de Febrero 210 bis (S2000EZP), Rosario, Argentina

[d]Universidad Nacional de Rosario, Facultad de Ciencias Exactas, Ingeniería y Agrimensura, Av. Pellegrini 250 (S2000BTP), Rosario, Argentina.

Corresponding author: Carmarán, Cecilia Cristina.

Tel/Fax + (54) 11 5285-8592

e-mail: carmaran@bg.fcen.uba.ar

Co- Corresponding author: Andres Ezequiel Dolinko

Tel/Fax + (54) 11 5285-8592

e-mail: adolinko@df.uba.ar

## Abstract

Currently, the detection of endophytic fungi is determined mostly by invasive methods, including direct isolation of fungal organisms from plant tissue in growth media, molecular detection of endophytic fungi DNA from plant material by PCR, or evaluation under microscopy techniques. In this work we explore the potential of laser biospeckle activity (LBSA) to be used for the detection of endophytic colonization of leaves of a promising energy crop, *Jatropha curcas* L. We compared the laser biospeckle activity of endophyte infected and uninfected *J. curcas* leaves. The differences between blade and veins (including midrib) of the studied leaves was validated and growth parameters of the studied plants were also analyzed using the normalized weigthed generalized differences coefficient (nWGD). The obtained results showed a relationship between the endophytic burden of leaves and the LBS, suggesting that LSBA is a useful tools to detect endophytic colonization in situ. Also, the increased water movements inside leaves promoted by endophytic colonization could be explain by the obtained data.

## 1. Introduction

The study of biological interactions and their impact, in the different biological and ecosystemic level, has expanded our interpretation and understanding about the organisms and their functioning. In plants, biological interactions occurring between functionally distinct symbionts that provide different benefits, often result in the promotion of key processes benefiting plant growth and health (Stanton 2003). An example of this would be to provide biotic and abiotic stress resistance and tolerance, to enhance nutrient availability, and to degrade toxic substances (Li and Zhang 2015). Fungal endophytes are defined as fungi that live asymptomatically within healthy plant tissues for at least part of their life cycle. This group of fungi has been found in all plant species studied to date (Carroll 1988; Wilson 1995; Jumpponen and Trappe 1998; Rodriguez et al. 2009; Porras-Alfaro and Bayman 2011; Kato et al. 2017), and appear to be ubiquitous in plants in natural ecosystems. The roles of the endophytes in ecosystems are just starting to be elucidated and little is known about the nature of the interactions between woody plants and their foliar endophytes. However, some studies have shown that this group of fungi may modulate the plant development as well as ecology of plant communities (Herre et al. 2007; Mejía et al. 2008; Yuan et al. 2011). Special research interest is focused on the endophytes of leaf tissues due to the multiple processes in which they are involved and the interactions of the leaf mycobioma with its hosts (Lindow and Brandl 2003; Rodriguez et al. 2009; Persoh 2015; Pandey et al. 2016).

Our ability to visualize complex interactions in the microbiome helps us to understand the different roles that endophytes can play. The research about the potential benefits of endophytic fungi, such as protection against herbivory or resistance to stress, involve the study of endophyte-colonized plants (E+) that must be monitored and be distinguishable from non-colonized plants (E-). Detection of endophytes is typically achieved through several techniques, including direct isolation of fungi from plant tissue using growth media (culture dependent), molecular detection of endophytic fungi DNA from plant material by PCR (culture independent), or histological techniques for microscopy (Hyde and Soytong 2008; McKinnon et al. 2017). All these methods involve a destructive analysis of the sample and the assumption that the results are a reflection of the existing endophytic communities in all the tissue, in the part analyzed and in the rest, that for different reasons cannot be analyzed. In this context, the appliance of non-invasive techniques represents a great advantage and increases our ability to detect this type of organisms *in situ*.

The speckle effect occurs when coherent light, such as the one coming from a laser source, is scattered by a rough surface or transmitted through a medium with random variations in its refractive index. The final optical field is the result of the interference of the light scattered by the optical inhomogeneities randomly distributed in the medium. Consequently, the resulting light intensity distribution presents a granular structure and is known as speckle pattern. If the scattering medium shows some type of activity that changes the optical path of the interfering light e.g., micrometric movements or small variations of refractive index, the resulting speckle pattern changes in time, and the visual appearance is similar to that of a boiling liquid. This effect is called dynamic speckle. The study of the temporal evolution of the dynamic speckle is a widely known non-contact and non-destructive technique, used to characterize the parameters involved in the processes that generate the activity of the scattering media (Rabal and Braga Jr 2008). If the scattering media is a biological sample, the effect is known as biospeckle. In this case, the activity observed is the result of movement inside the living tissue and can be attributed to many processes, such as growth and cell division, cytoplasmic movement, biomechanical reactions or water-related activities (Zdunek et al. 2014; Ansari and Nirala 2015). The study of the biospeckle activity, also named as biospeckle laser or LBSA (laser biospeckle activity), has been widely applied for analyzing different processes, such as plant seed activity analysis, blood flow detection, visualization of tissue perfusion, burn scar perfusion, fruit maturation, and assessment of leaf vein system (Zhong et al. 2014; Zdunek et al. 2014; Ansari and Nirala 2015).

*Jatropha curcas* L. (Family: Euphorbiaceae), is a promising energy crop that is being extensively studied to develop bio-fuel technology from its seeds, as well as for other beneficial uses. Studies in progress on this species, suggest that the artificial inoculation of endophytes can improve certain interesting agronomic features of plants (D'Jonsiles et al. 2019). However, the techniques used today do not provide non-invasive tools that demonstrate if the inoculation was successful or not.

The aim of this study is to test the potential of laser biospeckle activity (LBSA) for the detection of foliar endophytes colonization. This approach is based on comparing the biospeckle activity of endophyte infected and uninfected *J. curcas* leaves. The differences between blade and veins of the studied leaves was validated and growth parameters of the studied plants were also analyzed.

## 2. Materials and methods

### *2.1 Plant Material*

Seeds of *J. curcas* L. were provided by Ing. Diego Wassner (Cátedra de Cultivos Industriales, Facultad de Agronomía de la Universidad de Buenos Aires, Argentina). They were collected from experimental plots of *J. curcas* located in Siete Palmas, Formosa Province, Argentina (25° 13' 21.04"S, 58° 17' 59.67"W).

### *2.2 Endophyte analyses*

In order to reach the proposed objective, different procedures were performed to obtain leaves of *J. curcas* with different status of endophytic colonization (high or low), to analyze foliar tissue to determine the percentage of colonization through culture studies, to evaluate biomass indicators and to study the LBSA on leaves of *J. curcas*.

For these procedures 120 seeds were pregerminated in plastic trays, and transferred to plastic pots, with commercial soil (Terra fertile S.A.) to obtain young plants.

The foliar endophytic colonization is a natural process, thus, it was necessary to perform the following experimental procedure to obtain plants with low or no colonization: ten plants were covered with transparent acetate in order to avoid the infection, and ten plants were maintained uncovered (Arnold and Herre 2003).

For the different procedures carried out hereafter, two leaves (or cotyledons, in the corresponding case) per plant were analyzed. Each leaf was cut into pieces of about 2 x 2 cm. In the case of leaves analyzed by LBSA, endophyte isolation was performed on pieces corresponding to the evaluated areas (Fig. 1). All pieces were sterilized using the following immersion sequence: 1 min in 70% ethanol, 2 min in a 4% bleach solution, and 30 s in 50% ethanol. Each piece was transferred to 90-mm Petri dishes containing MEA: Malt Extract (OXOID) 2%, Agar (Difco) 2% supplementing with chloramphenicol (Sigma) 1%. Plates were incubated at 24°C for six weeks and examined periodically. Outgrowing mycelia were isolated, purified, transferred onto slants containing MEA and stored at 4°C. Once the fungi were isolated into pure cultures, percentage of endophyte isolation were registered as follow:

%EI= (Total of endophyte isolations/ total of fragments analyzed per leaf) * 100

The plants were analyzed taking into account two categories: endophyte infected (E +) or endophyte free (E-), the latter included those which presented an infection lower than 25%.

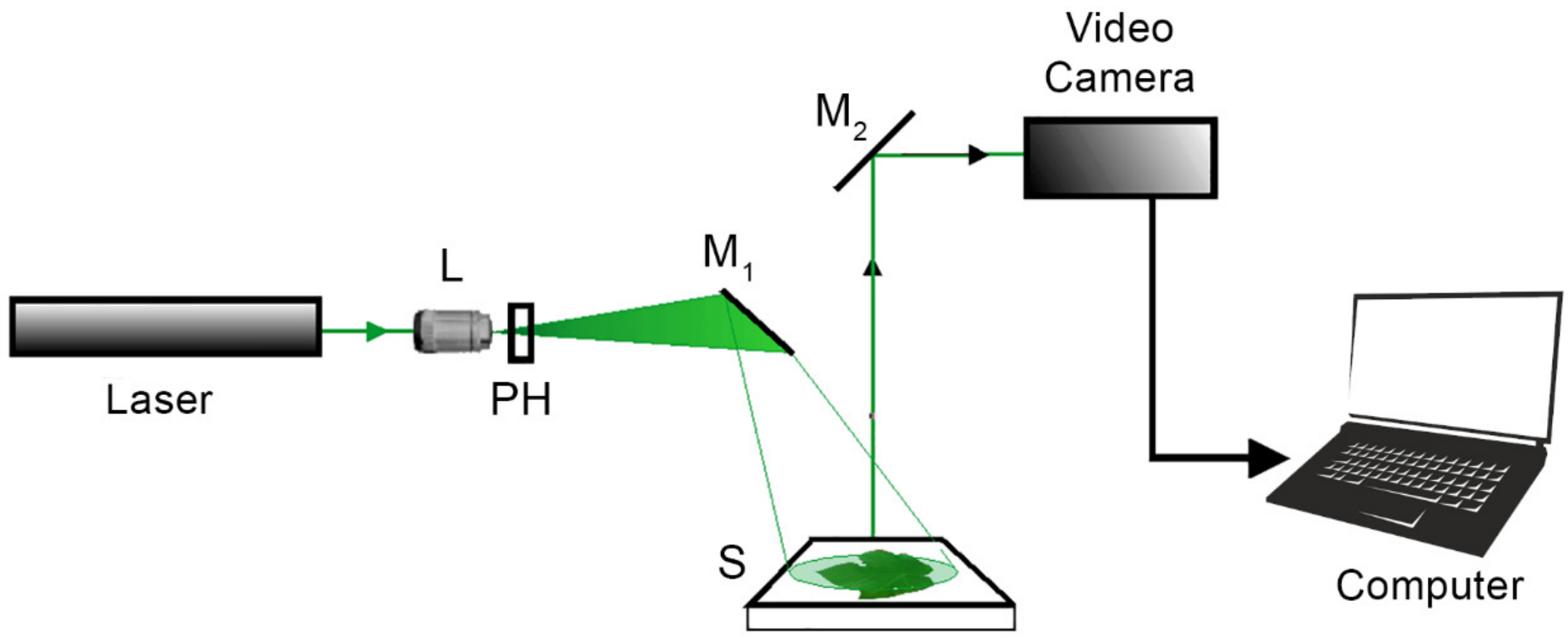

Fig. 1: Different procedures carried out in this work. Leaves of *Jatropha curcas* were first analyzed by laser biospeckle activity (LSBA), and then, the leaves areas studied were cut into pieces, sterilized and put in Petri dishes with nutritive culture medium for endophyte isolation.

*2.3 Endophyte leaf colonization*

In order to determine de adequate leaf age to analyze the endophytic colonization using LSBA, plants were grown during four months to keep the plants exposed long enough to the endophyte inoculum present in the environment. Endophyte isolations were carried out from cotyledons (that can persist even up to a month in the plant) and from leaves of 2, 4, 6, 8 or 10 days old. The findings (please see results for details) indicated that leaves of 10 days represented the adequate age to perform the following tests. Taking into account these preliminary results, twenty plants (10E+ and 10 E-) were grown during 4 months (from January to April 2017), under environmental conditions, and subsequently they were used to evaluate plant biomass and growing parameters. Additionally leaves of 10 days were used for the analyses carried out LSBA and endophyte isolation.

*2.4 Plant Biomass and growing parameters*

After four months, all plants were harvested and fresh weights (FW) were registered. The dry weights (DW) of shoots and roots were recorded after drying them in an oven at 60°C to constant weight. Root (RL), stem (SL) and total length (TL) were also measured.

*2.5 Biospeckle data analysis*

Among the 20 plants assigned to categories E+ or E- (10 E+ and 10 E-), 5 plants per category were chosen at random, and 2 leaves per plant were used for LBSA analyzes. Different methods have been developed to characterize biospeckle activity, and the choice of a specific approach depends on the nature of the phenomenon under study. In situations where the biospeckle activity is not uniform over the sample, image-processing based techniques such as Fujii method, generalized differences (GD) or weighted generalized differences (WGD) methods are more efficient to characterize the detected activity (Rabal and Braga Jr 2008; Ansari and Nirala 2015).

In order to analyze biospeckle activity in *J. curcas* leaves, we used the WGD coefficient $I'(i,j)$ defined as:

$$I'(i,j) = \sum_k \sum_l |I_k(i,j) - I_{k+l}(i,j)|\, p_l$$

(Arizaga et al. 2002), where $I'(i,j)$ the intensity of the point (or pixel) with $(i,j)$ in the k'th frame, the bars | | indicate absolute value and $p_l$ are user defined weights.

As every $I_k(i, j)$ value is subtracted from all other values in the same location, the result does not depend on the order of appearance of the $I_k(i, j)$ values. It results that the WGD is minimum (0) when all $I_k(i, j)$ are approximately equal, *i.e.* the activity is low or nonexistent. Conversely, it can be shown that the WGD value is a maximum when the $I_k(i, j)$ values in the histogram are evenly distributed near the bounds of the gray-level interval. This is so when the lower value occurs half the time, and the other half the higher one value appears. In our case, the gray-level interval is between 0 and 255. Thus, if the histogram of the intensity values of a pixel as a function of time shows two or more modes (i.e., maxima), the value of the WGD is greater when the modes are more separated.

If the $p_l$ weights are set to zero for all but the smaller $l$ values, relatively fast variations are enhanced with respect to slower ones. On the other hand, when processes are very slow with respect to the acquisition time of a frame and their variations are hidden by noise, it is more convenient to compare only frames that bear a significant time separation; that is, the $p_l$ weights are set to zero for small $l$ values and to 1 for some time interval where the process is expected to show significant variations (Arizaga et al. 2002).

### *2.6 Optical setup*

A diagram of the optical setup used in the experiments is shown in Fig. 2. In this study we used a green light He-Ne laser in order to minimize the light absorbed by the leaves and to be able to get the information of the leave surface. In this manner, we aim to minimize the interference with the leave photosystems. The green light from a Nd:YAG laser (Coherent model Compass 315M, 100 mW, lambda 532 nm) is expanded by means of the expansor lens (L) and directed to the specimen (S) by means of an illumination mirror ($M_1$). The light scattered by the specimen is directed to the high-speed video camera (Dalsa model CA-D6) externally driven by the frame grabber (Coreco Imaging PC-DIG) by means of an observation mirror ($M_2$). The specimen S (i.e. the plant leaf) lays horizontally on a supporting plate clamped to the optical bench. Additionally, the leaf is covered by a transparent acrylic plate to prevent spurious movements of the leaf. The camera was set to a frame rate of 25 frames per second for the experiments. Fig. 3a shows the assembly of the plants in the optical bench and Fig. 3b shows the optical set up during LBSA measurements.

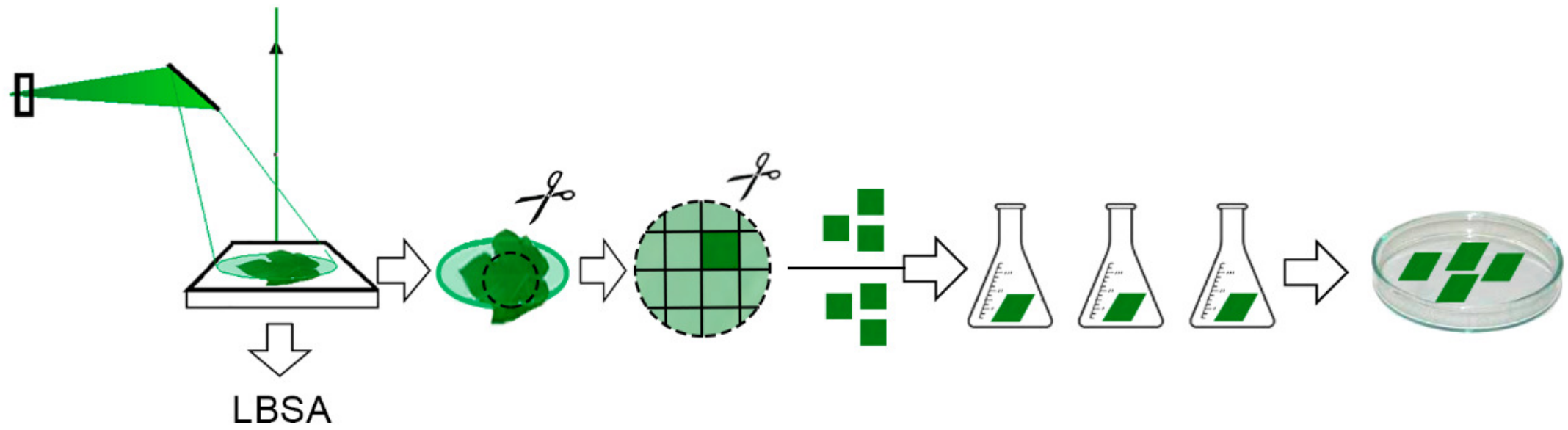

Fig. 2: Diagram of the experimental setup: Expansor lens (L), Pin hole (PH), Illumination mirror ($M_1$), Observation mirror ($M_2$), Specimen (S).

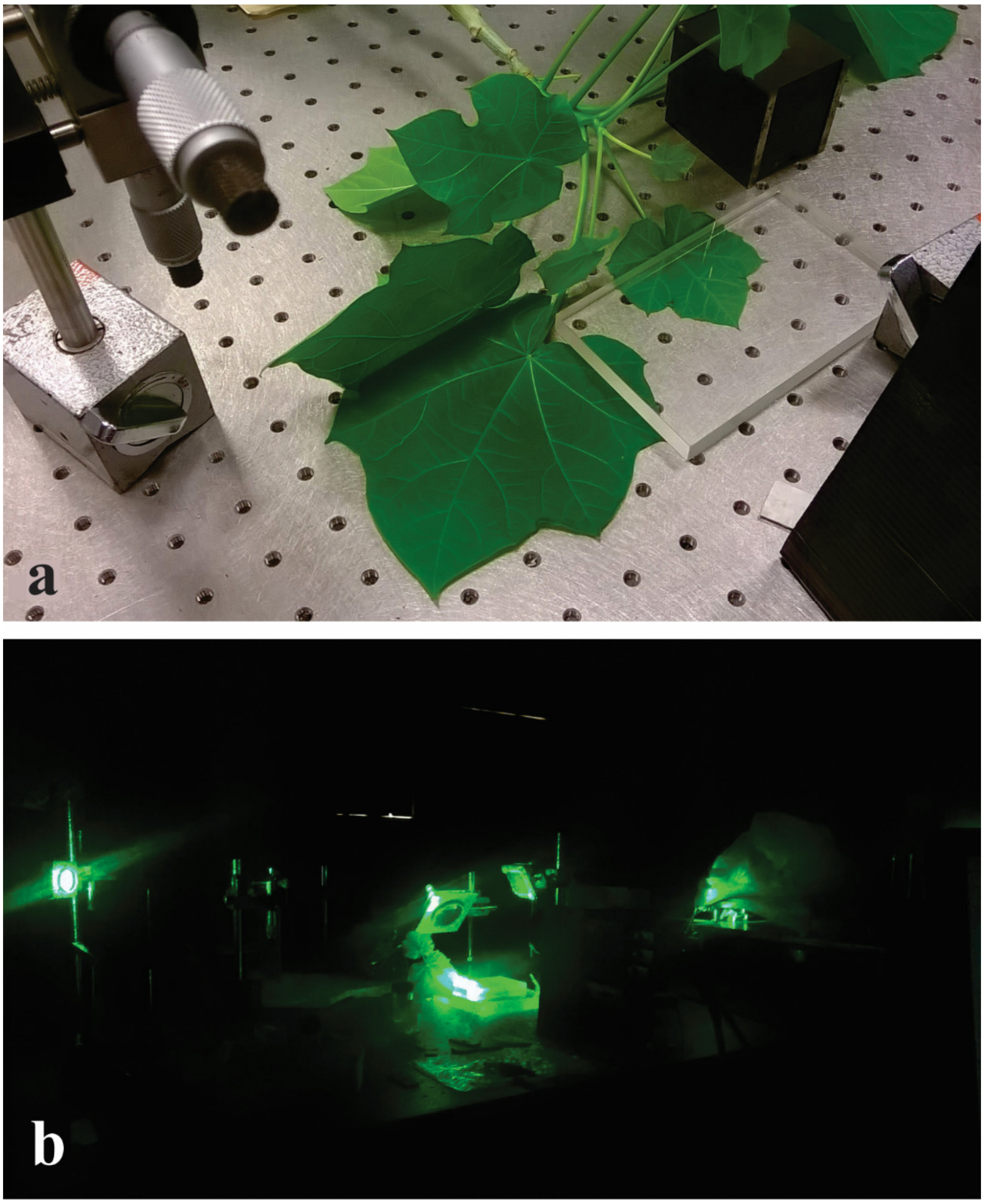

Fig. 3: Assembly of one *Jatropha curcas* leaf in the optical bench (a) and the optical set up (b) during the measurements.

*2.7 Image processing*

The sequences of dynamic speckle images where processed by evaluating the WGD coefficient. As explained above, this coefficient is evaluated for each pixel of the image sequence independently. We concentrated on the contrast of this coefficient between the region of the leaf blade and the regions of veins of the leaf, in order to cancel the possible influence of different variables as laser stability, color and reflectivity of each leaf and illumination, and also unknown surface variables of different leafs. To do this, the WGD image of each leaf was normalized to the maximum value of WGD in the image, obtaining a normalized WGD image (nWGD). In this manner, we obtain an image ranging from 0 to 1, where unity corresponds to the maximum value of WGD of each image. For this purpose, a histogram was elaborated with the values of frequencies reached in each measurement according to the obtained nWGD values. Then we compared the maximum frequency (MxF) and the corresponding nWGD values for all samples.

*2.8 Statistical analysis*

In order to compare differences between E+ and E- plants among all variables, a non-parametric Kruskal Wallis test was performed: nWGD (normalized weighted generalized differences); MxF (maximum frequency); FW (fresh weight); DW (dry weight); RL (root length); SL (stem length); TL (total length).

A multivariate analysis was applied to LBSA and growing parameters in 10 plants (5 E+/5 E-). Principal component analysis (PCA) was used to identify groups of plants showing similar patterns with respects to those variables, therefore, this analysis intended to extract the main trends, revealing the structure of the data. As the variables MxF and TL are directly obtained from nWGD and the sum of RL and SL, respectively, they were excluded from the analysis to avoid redundancy. When analyzing the original data set of quantitative variables, no potential outliers were found and all assumptions were met. Since the variables are expressed in different measurement scales, we compute a PCA on the correlation matrix to standardize the variables.

All analyses were performed with R version 3.4.3 2 (R Development Core Team, 2017).

**3. Results**

Results obtained from the study of endophytic colonization in function of leaf age (Table 1) showed that cotyledons and 2 days old leaves did not present endophytic infection, and 10 days old leaves carried bigger burden of endophytes than 4, 6 or 8 days old leaves. Therefore, subsequent analyzes were performed on 10 days old leaves from E+ and E- plants (see M&M) (Table 2).

Table 1: Percentage of endophyte isolation (%EI = (Number of endophyte isolation/Number of fragments analyzed)*100) of preliminary assays from cotyledons (Cot), and leaves of 2, 4, 6 and 10 days old of *Jatropha curcas*.

| **Leaf age** | **%EI** |
|---|---|
| **Cot** | 0,000 |
| **2** | 0,000 |
| **4** | 6,667 |
| **6** | 9,266 |
| **8** | 26,427 |
| **10** | 60,632 |

Table 2: Values of percentage of endophyte infection obtained from E- (free of endophytes) or E+ (infected with endophytes leaves of *Jatropha curcas.*

| **Leaves** | **% EI** |
|---|---|
| | 0 |
| | 0 |
| | 0 |
| **E-** | 0 |
| | 0 |
| | 13 |
| | 25 |
| | 25 |
| | 50 |
| | 62 |
| **E+** | 62 |
| | 75 |
| | 100 |
| | 100 |

Table 3 shows the values obtained for all the variables studied: nWGD (normalized weighted generalized differences); MxF (maximum frequency); FW (fresh weight); DW (dry weight); RL (root length); SL (stem length); TL (total length).

Differences between variables, analyzed through a Kruskal-Wallis test, are shown in Fig. 4. The results for biomass analysis showed that there are significant differences for DW ($H$= 7.62, $p$= 0.0057). For FW there were no significant differences ($H$=1.65; $p$= 0.19). The Kruskal-Wallis test for growing parameters revealed significant differences for RL ($H$= 9.19, $p$= 0.0024), but not for SL ($H$=0.05174, $p$=0.82) and TL ($H$= 1.97, $p$= 0.1598) (Fig. 4 a).

The LSBA was carried out on leaves of 10 days old from E- an E+ plants. The results for nWGD and MxF showed significant differences in both cases between E+ and E- leaves ($H$= 6.86, $p$= 0.0089; $H$= 6.81, $p$= 0.009 respectively) (Fig. 4 b and c respectively).

Table 3: Values obtained from E- (<25%of isolation), or E+ (>25% of isolation) *Jatropha curcas* leaves for all the variables studied: nWGD (normalized weigthed generalized differences); MxF (maximum frequency); FW (fresh weight); DW (dry weight); RL (root length); SL (stem length); TL (total length); %EI (percentage of endophyte infection). (*) adimensional values, (**) values in grs, (***) values in cm.

| **Leaves** | **(*) nWGD** | **(*) MxF** | **(**) FW** | **(**) DW** | **(***) RL** | **(***) SL** | **(***) TL** |
|---|---|---|---|---|---|---|---|
| **E- (<25%)** | 0.203 | 547 | 12,38 | 15,41 | 14 | 22 | 36 |
| | 0.221 | 634 | 8,59 | 12,31 | 15 | 17 | 32 |
| | 0.217 | 609 | 10,53 | 11,51 | 16 | 20 | 36 |
| | 0.242 | 498 | 15,47 | 12,38 | 14 | 26 | 40 |
| | 0.285 | 439 | 11,84 | 12,52 | 13 | 21 | 34 |
| | 0.221 | 557 | 5,9 | 1,09 | 13 | 24 | 37 |
| | 0.224 | 501 | 8,59 | 1,82 | 17 | 23 | 40 |
| **E+ (>25%)** | 0.281 | 399 | 3,04 | 0,65 | 8 | 15 | 23 |
| | 0.296 | 498 | 7,21 | 1,31 | 9 | 17 | 26 |
| | 0.314 | 403 | 12,81 | 2,76 | 8 | 24 | 32 |
| | 0.392 | 368 | 11,92 | 2,32 | 10 | 27 | 37 |
| | 0.371 | 373 | 5 | 1,17 | 11 | 19 | 30 |
| | 0.285 | 432 | 3,1 | 0,55 | 10 | 14 | 24 |
| | 0.342 | 353 | 6,33 | 1,35 | 7 | 21 | 28 |

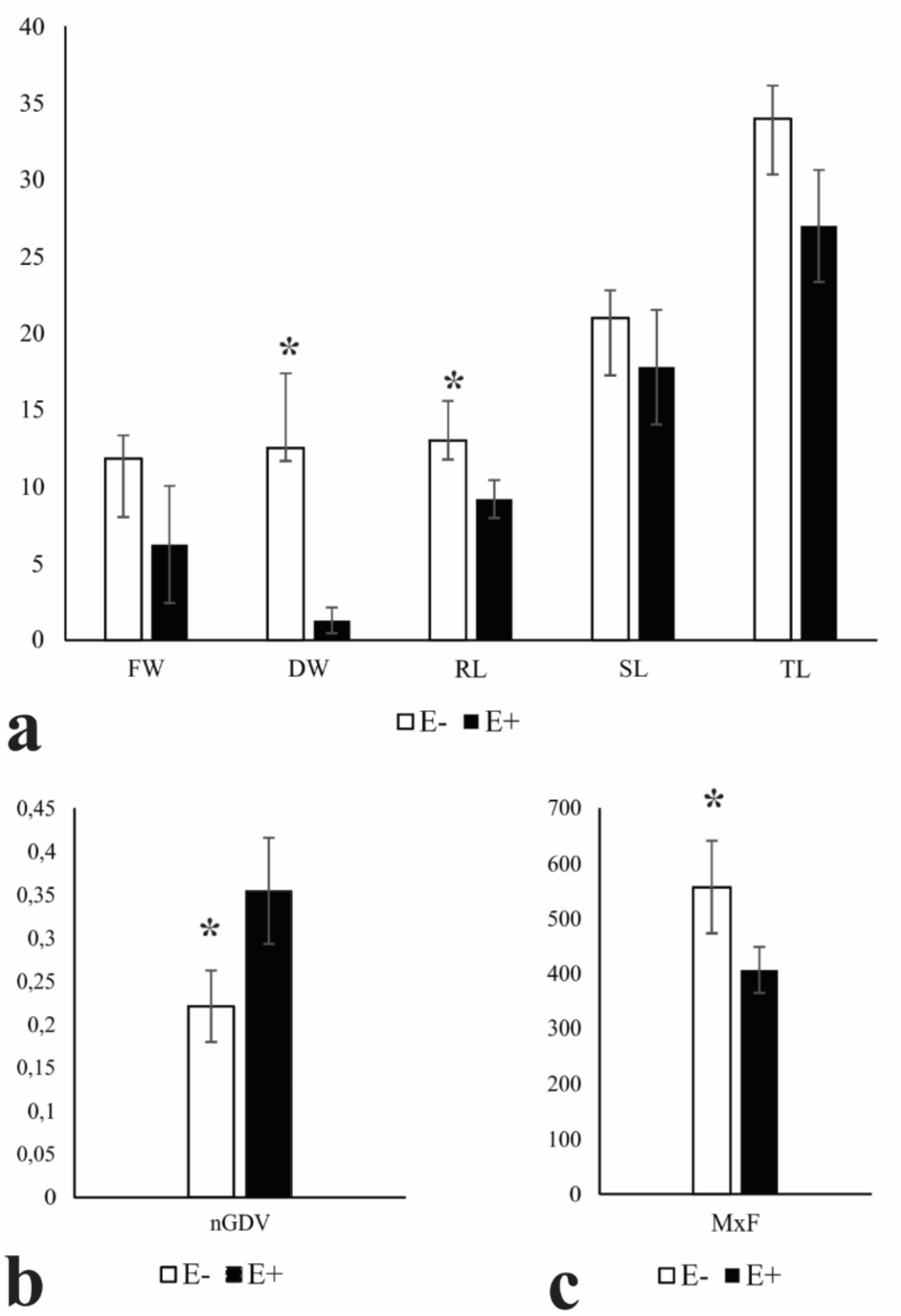

Fig. 4: Bar graphic showing Kruskal-Wallis test for E+ and E- plants with its standard deviation: a) differences for biomass variables (DW: dry weight, and FW: fresh weight) and for growing parameters (RL: root length, SL: stem length and TL: total length), b) and c) differences for LSBA(laser biospeckle activity) variables (nWGD: normalized weighted generalized differences, and MxF: maximum frequency respectively). Asterisks shows significant differences between E- and E+ plants for each variable (three graphics were performed because of the differences between scales)

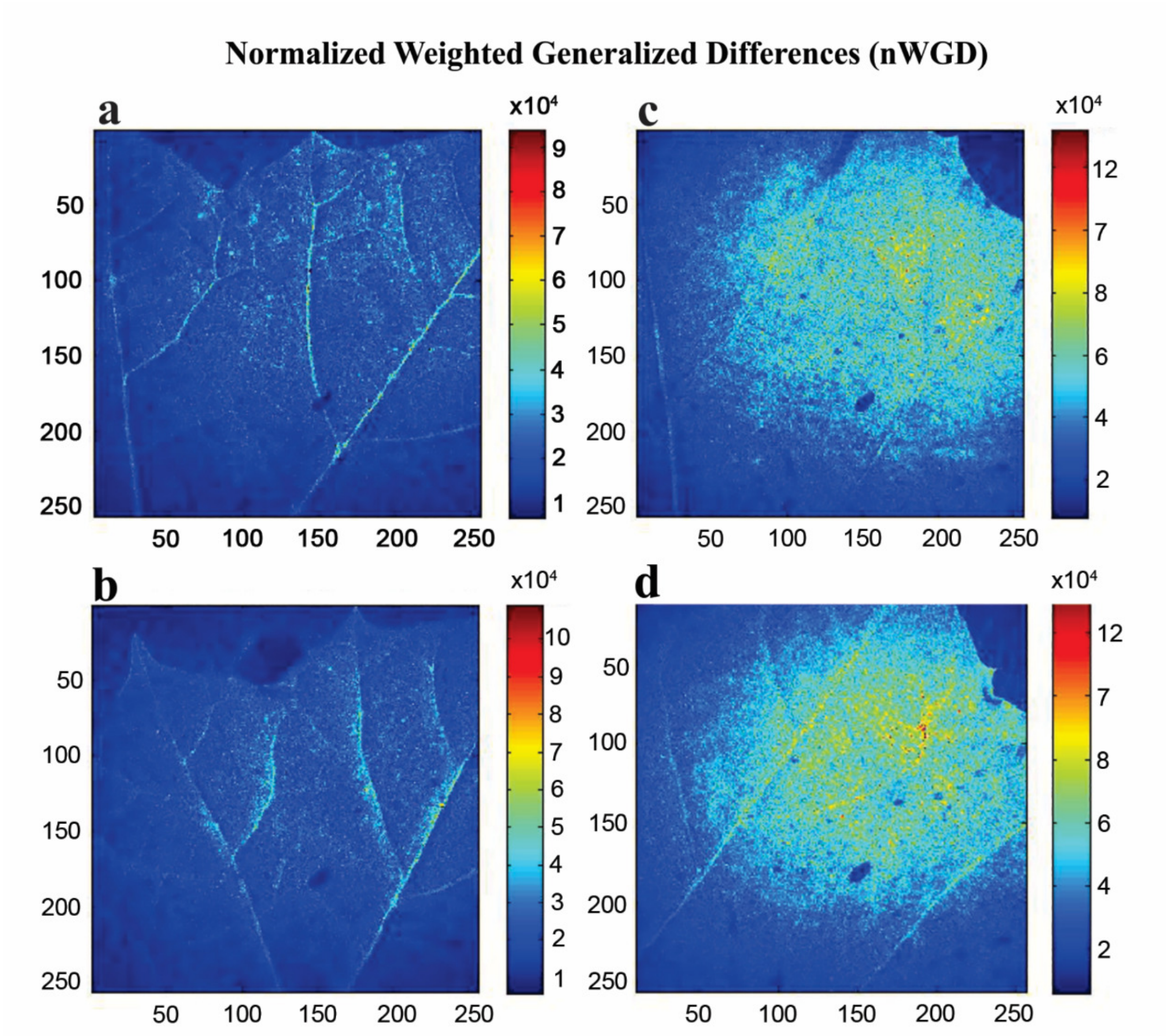

Fig. 5: Representative images of biospeckle activity obtained from plants analyzed, a) and b) represent E- leaves (free of endophyte), c) and d) represent E+ leaves (infected with endophytes). Note that the highest value are represented in yellow.

The nWGD images of the speckle image sequences of the evaluated leaves for E+ and E- shows observable differences. The nWGD for the E- cases studied shown lower values in the leaf blade than in the veins (Fig. 5 a-b). On the contrary, the nWGD for the E+ cases showed values in the same order of magnitude for the veins, and the region of the blade, as shown in Fig. 5 (c and d). The frequency of value of nWGD is represented by the histograms (Fig. 6), as a function of the nWGD range. The obtained data showed two different peaks, corresponding to E+ vs E- leaves.

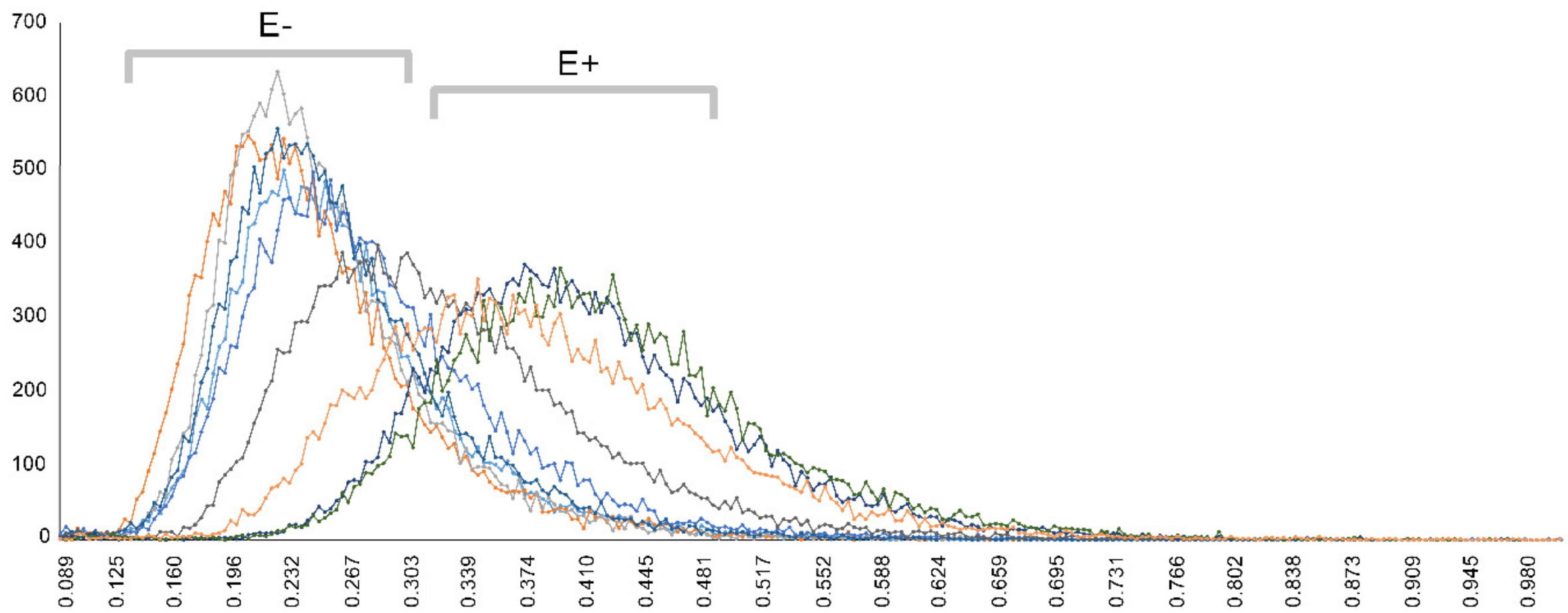

Fig. 6: Values of nWGD obtained after biospeckle data analysis. The histogram shows that lower values of nWGD (normalized weighted generalized differences) with high values of MxF (maximum frequency) correspond to E- leaves, while highest values of nWGD with low values of MxF correspond to E+ plants.

As a result of the PCA based on growth variables, including nWGD, the proportion of variance accounted for the first two axes was 85% (Axe I= 60%; Axe II =25%). Axis I was positively associated with nWGD and negatively associated with dry weight (DW), foliar weight (FW), length root (RL) and stem root (SL). Axis II was positively nWGD, SL and WW and this axis was negatively associated with RL .The plot shows that the plants clusters in two groups. The group located at the left was formed by plants 1–5, all E-, which displayed the highest values of DW, FW, RL and SL and the lowest values of nWGD. The group at the right included plants 6-10, all E+, which displayed the highest values of nWGD and the lowest values in the other variables (Fig. 7). The plant 6 (E+) was located at the positive end of the axe I, a little bit apart from the rest of the E+ plants. This plant showed the highest value of nWGD and higher values in FW, DW and SL, compared to the rest pf the E+ plants.

The Pearson coefficient ($r = 0.91$; $P = 0.001$) between the scores of axe I and endophyte percentage indicates a positive and significant association between these vectors.

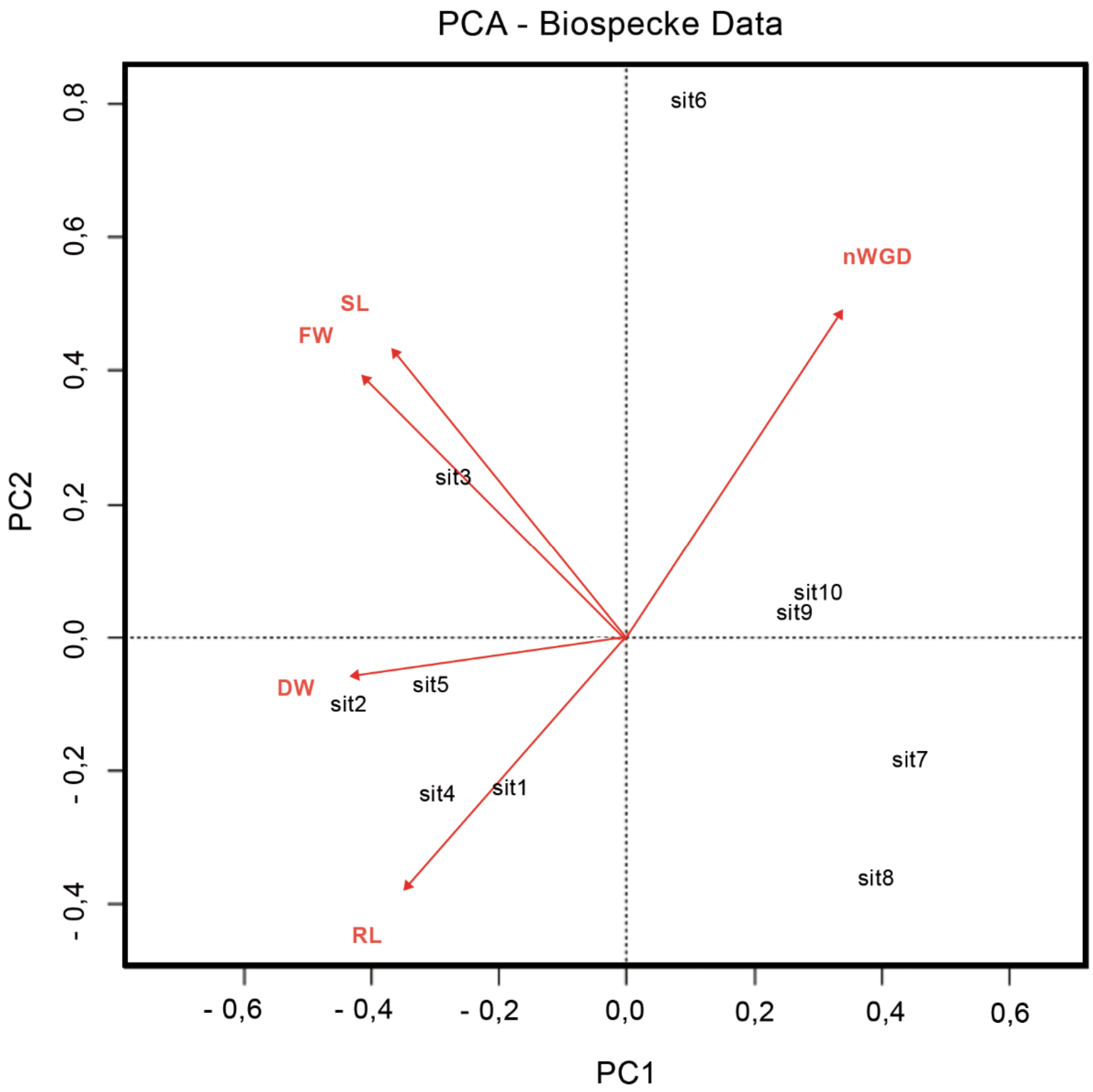

Fig. 7: PCA plot showing plants clusters in two groups. The group at the left (plants 1 to 5 all E-) with the highest values of DW (dry weight), FW (Fresh weight), RL (root length) and SL (stem length) and the lowest values of nWGD (normalized weighted generalized differences). The group at the right (plants 6 to 10, all E+) with the highest values of nWGD and lowest values for the rest of the variables.

**4. Discussion**

Considering that endophytes plays important roles in plant fitness, and traditional techniques to study them are most often destructive, here we apply a noninvasive technique to explore the LBSA in endophyte colonized and no colonized leaves of *J. curcas*. The results suggest that the Biospeckle activity could be used as a good indicator of endophytic colonization in leaves.

The LBSA is a technique that is being used in different fields of science. Braga Jr et al. (2007) presented results obtained from the application of biospeckle phenomena to detect fungi in beans (*Phaseolus vulgaris* L.). On the other hand, White et al. (2011) used this methodology for computing functional vascular density within a rodent dorsal window chamber model. For its part, Zhong et al. (2014) used this analysis for surface activity detection of attached leaves to monitor variation of leaf water status. Also Zdunek et al. (2014), reviewed the current stage of biospeckle technique and its applications, particularly for plant tissue evaluation. They concluded that the biospeckle method shows plenty of interesting nondestructive applications in agricultural crops. At present, in the area of quality evaluation, this method is still under development and has a chance to be commercially used as it is already utilized in medicine (Shvartsman and Fine 2003; Zdunek et al. 2014; Farraro et al. 2016, between others). Our study was based on the analyses of endophytes from leaves of 10 days old of *J. curcas*. We observed that there is a clear relationship between the endophytic burden of leaves and the biospeckle activity. Our data show that leaves with low endophyte infection (E-) have lower nWGD values, but have higher values of MxF, while leaves with high endophyte infection (E+) have higher values of nWGD but lower values of MxF.

Interestingly, multivariate analyses showed that E+ and E- plants conformed two distinctive groups. In particular, this could be related to biomass and growth parameters that differed significantly between those groups and that are correlated with endophyte infection. E+ and E- plants presented similar fresh weight, but E+ dry weight was significant lower than E- plants. These differences may be attributed to higher content of water in E+ plants. Furthermore, roots were significant longer in E+ plants than in E-. Both results suggest that E + plants would have a more efficient translocation of water than E- plants.

Plant–fungus associations, which are ubiquitous in plant communities, play important and varied roles in different plant physiological aspects (Hyde and Soytong 2008). Fungal pathogens associated with roots, vascular tissue and foliage may interfere with water uptake and transport, increase rates of foliar transpiration, and induce xylem embolism and tissue death (Agrios 1997). In contrast, rhizosphere mutualists such as ecto and arbuscular mycorrhizal fungi may benefit hosts by increasing surface area for water uptake, enhancing stomatal regulation of water loss, and increasing root hydraulic conductivity (Augé 2001; Lösch and Gansert 2002). Arnold and Engelbrecht (2007) used an experimental approach to show that endophyte colonization increases minimum leaf conductance, a measure of leaf water loss after maximal stomatal closure under drought stress, in the tropical tree *Theobroma cacao.* They showed that during maximum stomatal closure, leaves infected with a natural density and diversity of endophytes exhibit almost double the rates of water loss relative to uninfected leaves. In contrast, from the studies reviewed by Dastogeer and Wylie (2017), it is evident that endophyte colonization can significantly improve plant drought stress tolerance. These authors suggest that non-mycorrhizal fungi (for example foliar endophytes) may mediate the effects of water stress by adjusting, regulating or modifying plant physiological, biochemical and metabolic activities. In accordance with this, our data obtained for DW, and RL suggest that E + plants would have a more efficient translocation of water than E- plants, as mention previously. This is also coherent with the results obtained for LBSA analysis, where we observed a major activity in E+ leaves, and this response would be explained with water movements inside leaves.

An emergent property of the endophytic fungi as a result of the multiple interactions between plants and microorganisms, is its great phenotypic plasticity, that is, its ability to respond to various environmental signals. Our ability to visualize complex interactions in the microbiome helps us to understand the different roles that endophytes can play (Porras-Alfaro and Bayman 2011). To count with fast and noninvasive methods of detection of these organisms, such as the one we applied in this work, and to encourage research in this regard can be very useful in the area of agriculture.

## 5. Conclusions

LBSA is a promising noninvasive technique that is actually being used with medical purpose, and it is seen as a technique with great potential in agronomic areas. The present work represents a first approach for the development of a new tool to detect the status of endophytic colonization in a crop of interest, mediated by nondestructive and faster technique.

**Acknowledgment**

This study was supported by the National Council for Scientific and Technological Research (CONICET Argentina) PIP 11220150100956, PIP 11220110100971, and University of Buenos Aires UBACyT 20020150100067B

We thank the Instituto de Física de Rosario for providing us their facilities for optical measurements of this work, and Mariana Valente, for helping us with the editing of images.